\documentclass[12pt]{article}
\usepackage{sw20aip}

\input{tcilatex}
\begin{document}

\[
\text{Proposal for }in\text{ }situ\text{ Enhancement of Electron Spin
Polarization in Semiconductors.}
\]
\begin{equation}
\text{ H. Suhl}
\end{equation}
\[
Physics\text{ }Department\text{, }UniversityofCalifornia\text{, }SanDiego%
\text{, }9500\text{ }GilmanDrive\text{, }LaJolla\text{, }CA\text{ }92093%
\text{ }
\]

\medskip 

An extension of the original Overhauser effect to a more general
nonequilibrium state was proposed by G. Feher, and demonstrated by Clark and
Feher some forty years ago. It is suggested here that it might be possible
to produce excess electron spin polarization by allowing the role of the
nuclei to be played by other magnetic entities, such as paramagnetic
impurities or adjacent magnetically ordered structures.

\medskip 

\underline{I. Background:} Plans to utilize the spin degree of freedom of
electrons (rather than only their charge) in the construction of
semiconductor devices depend on the creation of a degree of spin
polarization well in excess of the very small net alignment available in
ordinary magnetic fields, especially at room temperature. An appealing way
to achieve large polarization is to inject electrons from the majority Fermi
sea in a ferromagnetic metal into the semiconductor. Giant magnetoresistance
heterostructures, involving injection of the polarized electrons into, and
transmission through, a nonmagnetic metal, raise hopes that such injection,
and subsequent transport, will be possible in semiconductors also. So far,
only a few successes have been claimed. These either involve cryogenic
temperatures, or else use semiconducting material not favored in applications%
$^{1}$. Also, in a certain semiconductor, anomalously high $g-$ values due
to band structure effects have been reported$^{2},$obviating the need for
polarized injection, but, again, very low temperatures are needed. Some
success has been reported involving optical methods$,$ and probably
reflection (rather than injection) from, a ferromagnet$^{3}$ .

Here, a method is proposed that avoids the need for injection or reflection
altogether, and should function at room temperature. It is based on a kind
of inversion of a generalized Overhauser effect proposed in 1959 by G. Feher$%
^{4},$ and subsequently realized experimentally by W. G. Clark and G. Feher$%
^{5}.$ The original Overhauser effect described greatly enhanced nuclear
spin polarization resulting from strong excitation of the paramagnetic
resonance of electrons in hyperfine interaction with the nuclei. The major
insight achieved by Feher$^{4}$ was that the crucial feature of the
Overhauser effect was not the microwave excitation of the electrons, but
simply their non-equilibrium distribution , no matter how produced. In the
Clark-Feher experiment, the electrons are thrown out of equilibrium by an
electric field applied to the semiconductor, ( indium antimonide), heating
the electrons to a temperature $T_{H}$, say. The hyperfine coupling of such
a 'hot' electrons to the In$^{115}$ nuclei causes a simultaneous spin flip
of the electronic and nuclear spins. Single spin flips of the electron by
spin-orbit coupling to the lattice also occur, but may be ignored initially.
Single spin flips of the nuclei lead to very long spin lifetimes and may be
ignored altogether. Let $A_{+},A_{-}$ denote the concentrations of upspin
and downspin hot electrons, and $B_{+},B_{-}$ the number of upspin and
downspin nuclei respectively. Then (since single spin flip processes are
ignored for now), the master equation for $A_{+}$ reads, in the steady state 
\begin{equation}
W_{-+\Rightarrow +-}A_{-}B_{+}-W_{+-\Rightarrow -+}A_{+}B_{-}=0  \label{1}
\end{equation}
where $W_{-+\Rightarrow +-}$ is the rate of simultaneous spin reversal of
electron and nucleus, the former from down to up, the latter from up to
down, and similarly for $W_{+-\Rightarrow -+}.$ Since this mutual spin flip
process, which does $not$ conserve energy, is powered by the hot electrons,
the ratio of the two $W$ 's has a value appropriate to detailed balance at
temperature $T_{H}$ : 
\begin{equation}
\frac{W_{-+\Rightarrow +-}}{W_{+-\Rightarrow -+}}=\exp \frac{2(\mu _{A}-\mu
_{B})H}{k_{b}T_{H}}\   \label{2}
\end{equation}
where $\mu _{A},\mu _{B}\ $are the magnetic moments of the electrons and
nuclei, and $H$ the applied magnetic field. In this open, non-equilibrium
system, the occupation numbers $A_{\pm },B_{\pm }$ will be given by
Boltzmann factors $e^{\pm \mu _{A}H/k_{b}T_{A}},e^{\pm \mu _{B}H/k_{b}T_{B}}$
with their own temperatures $T_{A}$ and $T_{B}.$Thus from equations (1) and
(2), 
\begin{eqnarray}
\frac{A_{+}}{A_{-}}\frac{B_{-}}{B_{+}} &=&e^{2\mu _{A}H/k_{b}T_{A}}e^{-2\mu
_{B}H/k_{b}T_{B}}  \label{3} \\
&=&e^{\frac{2(\mu _{A}-\mu _{B})H}{k_{b}T_{H}}}  \nonumber
\end{eqnarray}
whence 
\begin{equation}
\frac{\mu _{B}}{T_{B}}=\ \frac{\mu _{A}}{T_{A}}\ -\frac{\mu _{A}-\mu _{B}}{%
T_{H}}\   \label{4}
\end{equation}
This shows that for thermal energies of the hot electron far in excess of
the magnetic energies, $\frac{\mu _{A,B}H}{k_{b}T_{A,B}}\ ,$ the quantity
determining the extent of nuclear polarization approaches $\frac{\mu _{A}H}{%
k_{b}T_{A}}\ ,$ which determines the much bigger electronic polarization. In
other words, $\frac{B_{+}}{B_{-}}\rightarrow \frac{A_{+}}{A_{-}}$ as $%
T_{H}\rightarrow \infty .$ (Usually, this result is written as a greatly
reduced effective nuclear temperature $T_{B}=\left( \frac{\mu _{B}}{\mu _{A}}%
\right) T_{A}).$ Note that this assumes that the electron variables are
'robust', with $T_{A}$ rigidly fixed. Equation (4) could equally well slave $%
T_{A}$ to a rigidly fixed nuclear temperature. This difficulty is resolved
by taking single flip processes into account (see next section)

\underline{II. This Proposal:} The essence of this proposal is to let the
role of the nuclei be assumed by magnetic entities (for example paramagnetic
impurities with effective magnetic moments $\mu _{B}\ $much larger than the
electronic magnetic moment $\mu _{A}$ . Then $\mu _{A}\rightarrow \mu
_{B}\left( \frac{T_{A}}{T_{B}}\right) $ as $T_{H}\rightarrow \infty .$ In as
much as in such a system $\frac{T_{A}}{T_{B}}$ might be of order one, the
electrons will have acquired the magnetic moment of the impurity.

When single flip processes of the electrons and of the 'impurities' are not
neglected, it is found that this result retains its validity if a certain
inequality is satisfied. Note that the total concentrations of the $A$ and $%
B $ species scale out of equation (1) which is homogeneous of degree 2.
Inclusion of single flip processes spoils the homogeneity and results in a
concentration dependence. Writing $A_{+}=A\cos ^{2}\theta _{A},$ $%
A_{-}=A\sin ^{2}\theta _{A};B_{+}=B\cos ^{2}\theta _{B},$ $B_{-}=B\sin
^{2}\theta _{B}$ , with total concentrations $A,B,$the steady state master
equations for $A_{+},B_{+}$ now read 
\begin{eqnarray}
U+Aw_{+\rightarrow -}^{A}\cos ^{2}\theta _{A}-Aw_{-\rightarrow +}^{A}\sin
^{2}\theta _{A} &=&0  \label{5} \\
-U+Bw_{+\rightarrow -}^{B}\cos ^{2}\theta _{B}-Bw_{-\rightarrow +}^{B}\sin
^{2}\theta _{B} &=&0  \nonumber
\end{eqnarray}
where 
\begin{equation}
U=AB\left( W_{+-\Rightarrow -+}\cos ^{2}\theta _{A}\sin ^{2}\theta
_{B}-W_{-+\Rightarrow +-}\sin ^{2}\theta _{A}\cos ^{2}\theta _{B}\right)
\label{6}
\end{equation}
Here, $w_{+\rightarrow -}^{A}=w^{A}\exp (-\mu _{A}H/k_{b}T_{A}),$ $%
w_{-\rightarrow +}^{A}=w^{A}\exp (\mu _{A}H/k_{b}T_{B})$ are the single flip
rates for the electrons, and similarly for the 'impurities'. These flips are
powered by lattice vibrations via spin-orbit coupling, so that both $%
T_{A}\&T_{B}$ are presumably of order of the lattice temperature. Although
equations (5) and (6) can be reduced to a single quadratic, (for cos2$\theta
_{A},$ for example), the coefficients are very involved. However, there
appears to be one particularly simple solution for which $\theta _{A}=\theta
_{B}=\theta .$ If $U\neq 0,$ this solution, according to equations (5), must
satisfy 
\begin{equation}
Aw_{+\rightarrow -}^{A}\cos ^{2}\theta -Aw_{-\rightarrow +}^{A}\sin
^{2}\theta =-(Bw_{+\rightarrow -}^{B}\cos ^{2}\theta -Bw_{-\rightarrow
+}^{B}\sin ^{2}\theta )  \label{7}
\end{equation}
or 
\begin{eqnarray}
\cot ^{2}\theta &=&\frac{Aw_{-\rightarrow +}^{A}+Bw_{-\rightarrow +}^{B}}{%
Aw_{+\rightarrow -}^{A}+Bw_{+\rightarrow -}^{B}}  \label{8} \\
&=&\frac{1+\frac{Bw^{B}}{Aw^{A}}e^{(\mu _{B}/T_{B}\ -\mu _{A}/T_{A})H/k_{b}}%
}{1+\frac{Bw^{B}}{Aw^{A}}e^{-(\mu _{B}/T_{B}\ -\mu _{A}/T_{A})H/k_{b}}}
\end{eqnarray}
This is consistent with the earlier , concentration independent result $\cot
^{2}\theta =e^{2\mu _{B}/k_{b}T_{B}}$ for very large $T_{H}$, provided $%
\frac{Bw^{B}}{Aw^{A}}e^{-(\mu _{B}/T_{B}\ -\mu _{A}/T_{A})H/k_{b}}$ is much
greater than 1, and $\mu _{A}/T_{A}<<\mu _{B}/T_{B}$. (Note that, if $T_{H}$
were allowed to go to infinity,i.e. $U\rightarrow 0,$ at the beginning of
the calculation, this solution would fail.). An improved solution may be
obtained by writing $\theta _{A}=\theta +\delta _{A}$, $\theta _{B}=\theta
+\delta _{B}$ in equation (5), in the definition (6) for $U,$ and expanding
to first order in the $\delta $ 's, resulting in two first order linear
simultaneous equations for $\delta _{A}$ and $\delta _{B}$.

\underline{III. Possible Implementation.} In the above, the 'impurity' was
characterized as a simple magnetic moment $\mu _{B}$ and its Zeemann energy
in a magnetic field. To significantly enhance the electron spin
polarization, the implanted impurity must have a large spin and/or an
anomalously large $g-$ factor. In a magnetic field of 1 T, the electron
polarization ($A_{+}/A_{-})-1$ in the absence of the impurity would only be
0.6\% at room temperature. If the implant has a spin of 2, this figure would
be increased to about 2.4\%. (Although the analysis in section II was
phrased in terms of an impurity with spin 1/2, the results are easily shown
to hold for larger spins also, as long as the levels are equispaced). A
further increase could come if spin-orbit coupling to the crystal lattice
results in a large axial anisotropy energy for that implanted ion. This
might conceivably amount to an additional effective field of one Tessler,
giving 4.8\% excess polarization. However, much better results can be
obtained if the 'impurity' is replaced by be any magnetic structure with a
lowest magnetic excitation energy far in excess of any readily accessible
Zeemann energy..One promising case would be a pair of ions with spins
coupled by anisotropic exchange energy (isotropic exchange does not work,
since it commutes with the coupling $\vec{s}\cdot \left( \vec{S}_{1}+\vec{S}%
_{2}\right) $ to the conduction electron spin). For example, an exchange $%
J(S_{1x}S_{sx}+$ $S_{1y}S_{sy})$ would give an energy gap of order $J,$
commonly of order of several hundred cm$^{-1}.$ If $\mu _{B}H$ in the
foregoing results is replaced by this energy, it would give almost 100\%
electron spin polarization. (However, there may be a serious problem here:
energy conservation obviously requires that the electrons are hot enough to
deliver this kind of energy in the mutual spin flip process. Assuming a
mobility of 1000cm/sec/volt/cm, with the translational velocity acquired
from the electric field totally randomized, an energy gap of 100 cm$^{-1}$%
would require an electric field of about 10$^{4}$ volts/cm. Clark and Feher
in their experiment noted that a field of only 150 volts/cm already led to
breakdown, probably by impact ionization of donor ions.) Finally,
implantation may be avoided altogether by building a heterostructure
consisting of a thin semiconducting film sandwiched between two
antiferromagnetic insulators. If the anisotropy and exchange energies of the
latter are $J$ and $K$ respectively, their lowest excitation energy is of
order $\sqrt{JK},$again far above Zeemann energy in a commonly used magnetic
field. A full analysis requires allowing for position dependence of $A_{\pm
} $ and $B_{\pm }$ and excitation of the antiferromagnetic film along the
structure. This will be presented in a future calculation.

\medskip

References:

\medskip

1. P.R. Hammar and Mark Johnson, Phys. Rev. Letters 88. 066806, 2002

2. R. Fiederling, Nature, 40, 787, 1999

3. R.J. Epstein et al., Phys. Rev. B 65, 121202, 2002

4. G. Feher, Phys. Rev. Letters 3, 135, 1959

5 W.G. Clark and G. Feher, Phys. Rev. Letters, 10, 134, 1963

\end{document}